# DESIGN: ONE, BUT IN DIFFERENT FORMS

*Willemien Visser*

*INRIA Paris - Rocquencourt; CNRS LTCI; TELECOM ParisTech*

*46, rue Barrault*

*75634 Paris Cedex 13*

*France*

*tel: + 33 (0)1 45 81 83 19*

*fax:+ 33 (0)1 45 65 95 15*

*email: Willemien.Visser@Telecom-ParisTech.fr*

**Abstract.** This overview paper defends an augmented cognitively oriented generic-design hypothesis: there are both significant similarities between the design activities implemented in different situations and crucial differences between these and other cognitive activities; yet, characteristics of a design situation (related to the design process, the designers, and the artefact) introduce specificities in the corresponding cognitive activities and structures that are used, and in the resulting designs. We thus augment the classical generic-design hypothesis with that of *different forms of designing*. We review the data available in the cognitive design research literature and propose a series of candidates underlying such forms of design, outlining a number of directions requiring further elaboration.

This paper is a first step in an endeavour to assert an *augmented generic-design hypothesis* (which concluded our book, *The Cognitive Artifacts of Designing*, Visser, 2006b): *analysed from a cognitive viewpoint, design has specific characteristics that distinguish it from other cognitive activities, but also takes on different forms depending on the main dimensions of the design situation*. Examination of this hypothesis, which is the object of this paper, may have consequences for both theory and practice in the domain of design. Support for the hypothesis may have consequences for design environments, assistance, and education. It may, for example, guide the development of modalities for supporting designers when they are involved in the construction of representations or in the management of constraints and criteria. Given the mostly dispersed and anecdotal discussion of the different components that make up the hypothesis, our aim here is to articulate them in an overview paper.

Reviewing various empirical studies of activities "as diverse as software design, architectural design, naming and letter-writing," Thomas and Carroll (1979/1984) stated that these different design activities "appear to have much in common" (p. 234). A number of authors have defended that, compared to other professionals, designers have specific forms of knowledge (e.g., Cross, 2001b; 2002b ). Combining the positions underlying these two claims, Goel and Pirolli (1989; 1992) proposed the notion "generic design." Still other studies focus on the differences between design in different domains, examining a third aspect in this analysis concerning the nature of design (e.g., Akin, 2001; Purcell & Gero, 1996). In this paper we review and discuss these different aspects of design, focussing on the third one, whose discussion seems the least organised in the design literature. The position defended in this paper is the following: there are both (1) significant similarities between the design activities implemented in different situations and (2) significant differences between design and other cognitive activities; yet, (3) characteristics of a design situation (i.e., characteristics related to the design process, the designers, and the artefact) introduce specificities in the corresponding cognitive activities and structures that are used, and in the resulting designs. We thus augment the generic-design hypothesis (1 and 2) with that of different *forms* of designing (3).

The augmented generic-design hypothesis thus connects three different positions with respect to design and nondesign activities that have been espoused, more or less explicitly, by different authors in the domain of cognitive design research. Given the implicitness to this respect that is present in many papers on design, the review and discussion proposed in this paper seem useful. Generally, papers are concerned with only one position, sometimes two (the generic-design hypothesis), but the three have rarely been articulated together (but see Akin, 2001, discussed below). In addition, corroboration of the generic-design hypothesis is nearly exclusively grounded in Goel and Pirolli's (1989; 1992) work. Except for these authors' research, this double-sided hypothesis has received little substantiation through comparative cognitive analyses.

We qualify our hypothesis as "cognitive," because in the literature the term "generic design" most often is used in other than cognitive acceptations. The notion is used in the domains of software engineering (cf. Gamma's design patterns), AI and knowledge acquisition (e.g., KADS and successor work), based, for example, on Chandrasekaran's (1983) "generic tasks," or notions such as "generic design methods" and other "generic design agents" (see, e.g., Warfield, 1994). All these references are normatively based approaches to design, not concerned with the cognitive validity of the proposed units, be they design patterns, tasks, or methods. The present text focuses on cognitively oriented analyses of design activities.

*Outline of the paper*. This introduction presents our augmented generic-design hypothesis in the context of Goel and Pirolli's (1992) generic-design hypothesis and the view of design's domain independence defended by several other authors. Sections 1 and 2 briefly discuss the two constituents of the generic-design hypothesis, that is, the existence of commonalities between designing in different situations (section 1) and differences between designing and nondesigning (section 2). In the main section of this paper, that is, section 3, we discuss the third constituent of our hypothesis (that is, design takes different forms depending on characteristics of the design situation); we do so through an examination of candidate variables underlying such forms of design, outlining a number of directions for further elaboration. In section 4, the Conclusion, we will discuss the augmented generic-design hypothesis and complete it with a fourth constituent.

*The generic-design hypothesis*. Goel and Pirolli (1992) formulated their "intuitions about generic design" as a hypothesis that combined two assumptions: "problem spaces exhibit major invariants across design problem-solving situations and major variants across design and nondesign problem-solving situations" (p. 399).

According to Goel (1994), the authors aim to "motivate the notion of generic design within information-processing theory" (p. 53), that is, within the symbolic information processing framework that Newell and Simon (1972) developed in order to analyse problem solving from a cognitive viewpoint.

Goel and Pirolli (1989; 1992) seem to have made a strong case for the generic-design hypothesis. Their study may, however, be criticised on two points. On the one hand, certain flaws in the choice of nondesign tasks considerably weaken the authors' characterisation of design—which mainly depends on its contrast to nondesign (design is qualified as "X" by contrast to nondesign being "not X"). First, the nondesign tasks were brief, artificial games (that took 15 to 40 minutes). Second—something, moreover, noticed by the authors themselves—the study "purposefully took two points (ill-structured design tasks and well-structured game tasks) at the extremities of the spectrum of problem types" (Goel, 1994, p. 71). The author considers that, "given that [Goel and Pirolli] have found interesting differences, it would be instructive to… explore the intermittent points in the space" (Goel, 1994, p. 71), but this examination has not been conducted, as far as we know. On the other hand, the design tasks were examined in artificially restricted laboratory situations. The participants in Goel and Pirolli's (1989; 1992) study were professionals, but the design sessions, varying from 2 to 3 hours, "simulated the 'design sketch' exercises which are an integral part of the training program of many design disciplines" (Goel, 1994, p. 54)—tasks from which generalisation to design is not immediate, in our opinion. Such an approach is typical for the classical cognitive-psychology research on "problem solving" tasks. Since the beginning of our cognitive design research, we have been questioning the representativeness of such studies for professional design projects (in Visser, 1987b, for example, we identified specificities of professional design that are not observed in limited, artificially constructed design situations). In addition, we have come to question the appropriateness of the "problem solving" paradigm for the cognitive analysis of design (Visser, 2006b). The present paper focuses on cognitive design studies performed in real, professional work situations—even if we also review data from other experimental, less ecological research that is related to our topic. We will use, however, the notion "problem" and "problem solving" as authors use them, not always again questioning this view here.

*The domain independence of design*. Zimring and Craig (2001) present the "domain independence" of design as a notion similar to that of generic design. One may interpret domain independence, however, as only referring to design being invariant across domains, not necessarily to design differing significantly from other cognitive activities. It is in this more restricted sense that many design researchers and practitioners defend the idea of a domain-independent theory of design. Certain authors indeed defend such a position because of similarities observed between two or more domains of design.

Zimring and Craig (2001) consider that "common descriptions of design—that designing involves abductive reasoning, construction, ill-defined problem solving skills—…are not always sharp enough to both distinguish design from other types of problem solving and unite design across different design-related disciplines" (p. 125). The authors consider that the analysis and description of design in terms of "mid-level constructs" "may be more profitable in scaling research across disciplines" (p. 126). As examples of mid-level processes or types of reasoning, the authors present mental simulation, decision-making, and analogical reasoning.

During the 1995 "Design Dialogues: one" meeting, entitled "Universal Theory of Design: is a domain independent theory of design possible?"[1] the participants explored "the reasons for the apparent lack of progress in design research over [the decade 1985-1995] and in particular whether the search for an atemporal, acultural, domain independent theory of design [was] a reasonable or realistic goal." During this meeting, Cross stated, "a primary goal of the Design Research Society since its founding in the 1960s [had] been a domain independent theory of design within the context of a science of design" (from the Meeting report, see Note 1). In her Meeting report, McDonnell wrote, "on the question of whether theories, of whatever kind, can be domain independent, there was a… diversity of views.… some participants believing that some form of universal theory is possible ranged against those who argued for the incommensurability of different views of design or that the elimination of context necessary for a universal theory would result in an activity unrecognisable as design."

---

[1] "Design Dialogues: one" was "the first in an occasional series of discussion meetings on design theory sponsored by the Design Research Society," organised by McDonnell and Logan at University College London, on May 17, 1995. We quote McDonnell's meeting report, which until recently could be retrieved via internet, but is no longer accessible.

There has been much discussion in the design-research community around the relations between design and science (Sargent, 1994), some authors considering that a design science is to be developed (Hubka & Eder, 1987), others, such as Cross (2001b; 2002b), judging that the two are to be clearly distinguished. For Hubka and Eder (1987), "design science addresses the problem of determining and categorizing all regular phenomena of the systems to be designed, and of the design process" (p. 124). Cross (2002b) wishes to develop "'design as a discipline', based upon a 'science of design', not a 'design science'" (cf. also Simon, 1969/1999, characterising the "science of design" in his *Sciences of the Artificial*). For Cross (2002b), "design science implies an explicitly organised, rational and wholly systematic approach to design; not just the utilisation of scientific knowledge of artefacts, but design in some sense [as] a scientific activity itself.… Science of design refers to that body of work which attempts to improve our understanding of design through 'scientific' (i.e., systematic, reliable) methods of investigation. Let us be clear that a 'science of design' is not the same as a 'design science'. The study of design leaves open the interpretation of the nature of design." (see also Cross, 2001b)

"Domain" in the context of "domain independence" is generally equated with a "discipline (of practice)," such as engineering, architecture, computer science, or product design[2]. It may be used in a wider acceptation. Discussing domain-generality versus domain-specificity in cognition, Frensch and Buchner (1999, p. 142, quoted in Zimring & Craig, 2001, p. 126) define a domain as "anything that a given constraint can potentially be generalized to and from." In this paper, we will be concerned with design "situations" that can be characterised on three main dimensions, that is, the design process, the designers, and the artefact. The augmented generic-design hypothesis translates our claim that, if we do not "eliminate the context" of design (cf. McDonnell quoted previously), we may observe different forms of design in different design situations.

**1. Design is one: commonalities between designing in different situations**

From the early 1980s on, authors in the domain of design research have started to characterise design as a cognitive activity, highlighting the differences from design as it had been represented until then in prescriptive models underlying design methods (e.g., Pahl & Beitz, 1977/1996). An important reference for this new, more cognitively oriented approach to design has been Simon's (1969/1999) analysis of design in *The Sciences of the Artificial*. At the end of the 1990s, the following characterisation of design was prevailing in the domain of cognitive design research—even if authors may differ regarding certain characteristics (see hereunder *xi. Design activity is mostly opportunistically organised*). Concerning only two qualities, authors generally continue to adhere strictly to Simon's (1969/1999) position and analysis of design (see hereunder i and v). For most characteristics, authors have elaborated on Simon's characterisation, not so much contradicting him, as extending generally his analysis (see hereunder, especially, points iii, vi, vii, viii, ix, and x). For a last series, they have revised considerably Simon's position (see hereunder, especially, points ii, iv, and xi) (for a more detailed and more critical discussion of Simon's, 1969/1999, positions, see Visser, 2006b).

---

[2] We use the term "product design" where many authors use "industrial design," because we reserve "industrial design" in a more general acception, that is, for design perfomed in a professional, industrial situation.

(i) Design is a type of cognitive activity rather than a professional status. In 1969, Simon (1969/1999) states that "design" is not restricted to engineers, who are not the only professional designers. "Everyone designs who devises courses of action aimed at changing existing situations into preferred ones." (p. 111)

(ii) Design is a problem-solving activity. This is one of Simon's central stances with respect to design, based on the symbolic information-processing framework developed in Newell and Simon (1972). In addition, Simon qualifies design as an "ordinary" problem-solving activity, that is, a problem-solving activity for which no new and hitherto unknown problem-solving concepts or techniques are necessary. According to his "nothing special" position (presented for scientific thinking in Klahr & Simon, 2001, p. 76), "no qualitatively new components" need to be introduced in the classic general problem-solving mechanisms, in order to be able to handle design problems (Simon, 1973/1984, p. 197). No "special logic" is necessary (Simon, 1969/1999, p. 115)—even if Simon "admits" that standard logic is to be adapted to the search for alternative solution elements (p. 124). In recent years, we have started to amend Simon's (1969/1999) position: we have developed the idea that designing is more appropriately qualified as the construction of representations (Visser, 2006b, 2006c). From a formal viewpoint, design is certainly a "problem solving" activity: based on the design specifications, designers are rarely able to evoke from memory a pre-existing problem-solving procedure. Numerous studies have shown that, for many components of a design task, designers need to construct procedures in order to formulate a solution. However, qualifying design "simply" as problem solving is not very informative. In this paper, we cannot further detail these ideas (see Visser, 2006b).

(iii) Design problems are considered ill-defined (or "ill-structured" in Simon's, 1973/1984, terms): this design feature, noticed from the earliest cognitive design studies on (Eastman, 1969, 1970; Reitman, 1964; Thomas & Carroll, 1979/1984; Voss & Post, 1988), has been substantiated in many different kinds of studies since then, and continues to be considered as a specific characteristic of design (Akin, 2001; Michalek & Papalambros, 2002; Ormerod, 2005). Rittel and Webber (1973/1984) speak of "wicked" problems, which have no definitive formulation: each formulation corresponds to at least one solution (Buckingham Shum, 1997; Conklin, 2006) (for a discussion of the distinction between ill-defined and wicked problems, see Visser, 2006b, p. 142).

(iv) In his problem-solving approach to design, Simon (1969/1999; 1973/1984) distinguishes two stages in problem solving: problem structuring and problem solving. Analysis, synthesis, and evaluation are examples of another decomposition of design proposed by authors adopting, with more or less profound modifications, Simon's approach to design—or, more generally, Newell and Simon's (1972) approach (Akin, 1986a, 1986b; Baykan, 1996; Goel, 1994; Goel & Pirolli, 1992; Hamel, 1995; Lebahar, 1983). However, such stages can be distinguished only in theory as distinct activities: problem analysis and solution elaboration progress in parallel, rather than in separate, consecutive stages. Furthermore, designers constantly generate new task goals and redefine task constraints. Even if they are cognisant of prescriptive models distinguishing analysis and synthesis, designers do not follow them systematically (Akin, 1979/1984; Carroll & Rosson, 1985; Cross, 1984; Dasgupta, 1989; Visser, 1987a). Authors who analyse design problem solving in terms of "problem space" and "solution space" have proposed the notion of "co-evolution" of these two spaces (Dorst & Cross, 2001; Maher, Poon, & Boulanger, 1996; cf. also our idea of problem/solution pairs, Visser, 1991).

(v) Design is a "satisficing" activity: rather than to optimize, that is, to calculate the optimum value, or to choose the best solution among all possible solutions, designers "settle for the good enough" (Simon, 1971/1975, p. 1), accepting a satisfactory solution (Simon, 1987/1995, p. 246). As they have to decide without complete information, they have no other choice. This characteristic has been observed by various authors (Akin, 2001; Ball, Lambell, Reed, & Reid, 2001). According to Akin (2001), however, designers from different disciplines vary on this point: while architects indeed proceed to satisficing, engineering designers adopt more objective methods in their selection among possibilities and may proceed to optimisation.

(vi) Design generally involves complex problems that are rarely decomposable into independent subproblems. Of course, designers proceed to decomposition, in order to make their problems more manageable and easier to solve. In our view, Simon and many design researchers who follow him overestimate, however, the role of *systematic* problem decomposition, especially through balanced, stepwise refinement. In other than relatively routine design projects, designers rarely decompose in a systematic way (cf. our critique of Simon's "overestimating the role of systematic problem decomposition," Visser, 2006b, pp. 68-70). Moreover, one and the same design component often can be decomposed in different ways (Reitman, 1964, p. 296). Simon himself notes that the interdependencies among the subproblems resulting from problem decomposition "are likely to be neglected or underemphasized." "Such unwanted side effects accompany all design processes that are as complex as the architectural ones" that he considers in his text (Simon, 1973/1984, p. 191). According to Akin (2001), architects use idiosyncratic strategies to decompose a problem into subproblems and to integrate their solutions into a global solution afterwards, whereas in electronic hardware or mechanical design, the interaction between the parts are "theoretically determined." Notice that Simon (1973/1984, pp. 200-201) analyses complex systems such as social systems as "nearly decomposable" and that Goel (1995) considers the modules resulting from decomposition as "leaky."

(vii) Designers often tend to generate, at the very start of a project, a few simple objectives in order to create an initial solution kernel to which they then are sticking in what is going to become their global design solution. Such an initial solution kernel, which Darke (1979/1984) qualified as "primary generator," has been identified by many other authors and has received labels such as "kernel idea," "central concept," "early solution conjecture," "primary position," and "guiding theme" (Cross, 2001a, 2004b; Guindon, Krasner, & Curtis, 1987; Kant, 1985; Lawson, 1994; Rowe, 1987; Ullman, Dietterich, & Staufer, 1988).

      The ensuing process has been qualified as "position-driven" design, "early fixation," "premature commitment," "early crystallisation," or "solution fixation" (Ball, Evans, & Dennis, 1994; Cross, 2001a; Goel, 1995) (we will come back upon this characteristic, arguing that it requires inspection).

(viii) Rather than one solution, which would be "the" "correct" solution, design problems have several, acceptable solutions, which are more or less satisfying. This characteristic of design problems, related to their ill definedness and the satisficing character of designing, has been observed in many studies and domains, for example, architecture (Akin, 2001; Eastman, 1970), mechanical design (Frankenberger & Badke-Schaub, 1999), software design (Malhotra, Thomas, Carroll, & Miller, 1980), and traffic-signal setting (Bisseret, Figeac-Letang, & Falzon, 1988).

(ix) Design problems and solutions lack pre-existing, objective evaluation criteria (Bonnardel, 1991; Ullman et al., 1988). As evaluative references are forms of knowledge, designers' expertise in a domain influences how they use them (D'Astous, Détienne, Visser, & Robillard, 2004). Given that, in a collaborative design setting, designers may have different representations of their project, solution proposals are evaluated not only based on purely technical, "objective" evaluative criteria; they are also the object of negotiation, and the final agreement concerning a solution often results from compromises between designers (Martin, Détienne, & Lavigne, 2001). In addition, not only solution proposals, but also the evaluation criteria and procedures themselves undergo evaluation (D'Astous et al., 2004).

(x) Reuse of knowledge (from specific previous design projects) through analogical reasoning has been observed in many cognitive design studies as a central approach in design (Ball & Christensen, 2007; Ball, Ormerod, & Morley, 2004; Bhatta & Goel, 1997; Burkhardt, Détienne, & Wiedenbeck, 1997; Casakin & Goldschmidt, 1999; Détienne, 2002; Maiden, 1991; Sutcliffe & Maiden, 1991; Visser, 1995, 1996). Of course, this use of specific knowledge is combined with that of generic knowledge (especially, from design methodology, the application domain, and the technical domains that underlie the design project). Most examples of reuse concern software design (Détienne, 2002; Visser, 1987b), but we also observed it on product design (in the Delft study, Visser, 1995).

(xi) Design activity is mostly opportunistically organised: designers proceed in a non-systematic, multidirectional way (at moments top-down, at others bottom-up, at moments in breadth, at others in depth), formulating plans that are more or less local, at both high, abstract and low, concrete levels. The basis for such organisation is designers taking into consideration the data that they have at the time: specifically, the state of their design in progress, their representation of this design, the information at their disposal, and their knowledge (cf. the qualification of design as "situated").

This last point needs some discussion, because not all researchers share our conclusion that design is opportunistically organised (for a detailed discussion, see ch. 21.4 *The opportunistic organization of design: Decomposition and planning*, and especially its section "Discussion of our opportunistic-organization position," in Visser, 2006b, pp. 163-177). Especially Davies (1991) and Ball and Ormerod (1995) adopt other positions. According to Davies (1991), "expert programmers adopt a broadly top-down approach to the programming task, at least during its initial stages" (p. 186; see our discussion of expertise below in Section 3.2). Ball and Ormerod (1995) claim, "much of what has been described as opportunistic design behavior appears to reflect a mix of breadth-first and depth-first modes of solution development" (p. 131), even if design is also "subject to potentially diverging influences such as serendipitous events and design failures" (p. 145). Obviously, designers may proceed top-down and depth-first, or top-down and breadth-first—or bottom-up combined with depth-first or breadth-first. What we wish to emphasise is that (1) they often do so occasionally and locally, rather than systematically throughout the entire design process; (2) a top-down - bottom-up mix can take different forms, and even if a mix pattern has several occurrences—and thus gets a systematic character—these will generally be interspersed with other ways of proceeding, so that "top-down" and "bottom-up" are inappropriate as general qualifications of designers' activity; and, especially, (3) an occasional, local top-down - bottom-up and or breadth-first - depth-first mix are just some of the various forms in which opportunism can reveal itself in design.

As regards the "broadly top-down with opportunistic local episodes" (Davies, 1991) *versus* "opportunistic, with hierarchical episodes" (Visser, 1994a) issue, we follow Hayes-Roth and Hayes-Roth (1979, p. 307). These authors have proposed that the systematic refinement model be considered as a special case of the opportunistic model, which allows various organisational structures of an activity— rather than only one, or a mix of two structures. An opportunistically organised activity may have hierarchical episodes at a local level, but its global organisation is not hierarchical (Visser, 1994a).

With respect to the structured character of design organisation, opportunism proponents (Guindon et al., 1987; Kant, 1985; Ullman et al., 1988; Visser, 1987a; Voss, Greene, Post, & Penner, 1983) question the systematic implementation both of a depth-first (or breadth-first) and of a top-down refinement (or bottom-up) approach.

Notice that Goel (1995), who presents design as a quite systematic process, also remarks that "designers differ substantially in the path they take through [the design problem] space and how quickly or slowly they traverse its various phases" (p. 123). In addition, he notices that "problem structuring" (in his model, the first phase of design development) "occurs at the beginning of the task,… but may also recur periodically as needed" (p. 114).

These 11 qualifications, based on studies in different application domains, have contributed much to the development of the position that there are important commonalities between the implementations of design in different domains, that is, one of the two components of the generic-design hypothesis. Despite the more or less implicit adherence to this hypothesis in the design literature, there has been little systematic empirical research, however, to corroborate it—that is, apart from Goel and Pirolli's (1989; 1992) work. In the rest of this section, we present some rare studies concluding to the existence of more or less similar features between designing in different situations.

There is a series of early cognitive design studies conducted by Carroll and various colleagues in different design disciplines. In their review of this work, Thomas and Carroll (1979/1984) conclude that software design, architectural design, naming, and letter writing have many commonalities (as noticed in our introduction).

Bringing together observations gathered on product design (in the Delft study, Visser, 1995) and on software design (Visser, 1987b), we concluded that designers from these two disciplines proceeded to reuse.

Reymen et al. (2006) have performed empirical case studies in three design disciplines—architectural, software and mechanical design—in order to develop domain-independent design knowledge. The authors conclude that the supposed "important differences" between these design disciplines "concern mainly differences in terminology" (p. 151). One may notice, however, that the authors did not observe designers at work. They conducted interviews with designers concerning particular projects and analysed the documentation of these projects.

In his paper *How is a piece of software like a building? Toward general design theory and methods*, Gross (2003) advances the thesis that these two types of artefacts are alike on several dimensions: their size, level of complexity, lifetime, and degree to which their components are subject to change, the proportion of reusable components in their structure, the sanitary risks and safety concerns that particular uses or states of these artefacts may introduce, the type of their use or user, and the differences between their client and user. However, Gross (2003) does not refer to empirical work. We will come back to most of these dimensions below.

Notice that in the abovementioned studies, presented in order to show commonalities between designing in different situations, these different "situations" were always different domains of discipline—never different conditions of age, sex, expertise, or working conditions (e.g., process variables), for example.

**2. Design is different from nondesign**

The idea that design significantly differs from nondesign activities is stated explicitly less often than its counterpart in the generic-design hypothesis (that is, that design activities implemented in different situations are significantly similar). Cross (2001b; 2002b ) contends—as the underlying axiom of the design discipline he defends— that there are "forms of knowledge special to the competencies and abilities of a designer." Yet, it is not trivial to indicate what makes design specific.

*Expertise: design versus nondesign*. In his overview paper on design expertise, Cross (2004b) concludes that "expertise in design has some aspects that are significantly different from expertise in other fields" (p. 427). Referring to results from several empirical studies, he observes that the classical depth-first (novices) - breadth-first (experts) (or top-down - bottom-up) difference is not as systematically displayed in design as in other problem-solving tasks. The other main characteristics identified by Cross (2004b) are the following. A key feature of design expertise is "problem framing," that is, structuring and formulating the problem. Expert designers are solution-focused rather than problem-focused, especially in their particular domain of expertise (cf. also Lawson's, 1979/1984, results differentiating students from architecture and science on the solution-focused - problem-focused dimension). Expert designers frequently switch between different types of cognitive activity. Contrary to what is considered "good practice" in design methodology, they readily commit to an early solution concept that they elaborate, at all costs, patching it, if necessary: they do so instead of generating and examining a large number of alternative solution concepts, and of abandoning their solution concept when confronted with problems in its development (cf. point vii in Section 1).

Advocating that, from a cognitive viewpoint, design is definitely different from other activities, Goel and Pirolli (in Goel, 1994; Goel & Pirolli, 1989) quoted the examples of chess and medical diagnosis. These were, however, not the nondesign tasks examined by the authors in order to corroborate their generic-design hypothesis. To this aim, they compared protocols (collected by Newell & Simon, 1972, and published in their *Human Problem Solving*) concerning cryptarithmetic and the Moore-Anderson logic task with protocols the authors themselves had collected in three design disciplines, architecture, mechanical engineering, and instructional design (Goel, 1994; Goel & Pirolli, 1989). In their conclusion, the authors notice that, now that results have been obtained with such extreme well-structured tasks (cryptarithmetic and the Moore-Anderson logic task are two perfectly defined play problems), other activities need to be examined.

In *The Sciences of the Artificial*, Simon (1969/1999) presents as different the cognitive activities implemented in economics and in design. Analysing economic theories, he is very sensitive to the way in which economists idealise human rationality and neglect its limits. With respect to design, however, Simon seems to underestimate human cognitive limitations—something we illustrated in some detail in *The Cognitive Artifacts of Designing* (Visser, 2006b).

Several authors discuss differences between design and science (cf. also our introductory section). According to Archer (1979) "there exists a designerly way of thinking and communicating that is both different from scientific and scholarly ways of thinking and communicating, and as powerful as scientific and scholarly methods of enquiry when applied to its own kinds of problems" (p. 18). This idea of "designerly ways of knowing" is further developed by Cross (1982) (see also Cross, 2006).

Lawson (1979/1984) compared students from architecture and science (using artificial tasks supposed to represent architectural-design activities). He observed that the science students analysed their problems in order to discover their structure, whereas the design students generated "a sequence of high scoring solutions until one proved acceptable" (p. 218). Lawson's (1979/1984) conclusion has constituted one of the bases for distinguishing architects from scientists, encountered in papers qualifying architects as "solution-focused" and scientists as "problem-focused" (see also Kruger & Cross, 2006). Other studies seem to show, however, that a solution-focused approach is related to one's experience (Cross, 2004b; Lloyd & Scott, 1994).

In addition, various authors perceive elements of similitude between scientific and design activities. In his analysis of the structure of design processes, Dasgupta (1989), for example, considers design problem solving a special instance of scientific discovery. In *The Sciences of the Artificial*, Simon (1969/1999) establishes a correspondence between social design (social planning) and scientific discovery: they share a type of search, that is, a "search guided by only the most general heuristics of 'interestingness' or novelty" (p. 162). Cagan, Kotovsky and Simon (2001) point out the cognitive and computational similarities between the "seemingly disparate activities" of scientific discovery and inventive engineering design (p. 442). They notice that highly creative design activities are often labelled invention. The major conclusion of the authors' comparison is that, "at a deep level, the cognitive and computational processes that accomplish [design and discovery] are virtually identical" (Cagan et al., 2001, p. 463). These underlying cognitive activities are based on "problem solving, pattern recognition, analogical reasoning, and other cognitive knowledge retrieval mechanisms" (pp. 452-453). The authors thus defend, with respect to scientific-discovery and inventive engineering-design problems, their "nothing special position" (see point ii in Section 1). They establish, however, a "fundamental difference" between the two: "the *goal* of the process: Scientific explanation versus creation of a new artifact. . . . Design starts with a desired function and tries to synthesize a device that produces that function. Science starts with an existing function and tries to synthesize a mechanism that can plausibly accomplish or account for that function" (Cagan et al., 2001, p. 455). We do not share this reserve concerning the goals of the two activities: in our view, ultimately science (be it discovery or invention) is a design activity, the artefact aimed here being a theory.

In short, except with respect to expertise, there is little direct evidence for design differing significantly from nondesign. Indirect evidence might come from the previous section. Implicitly, our stance is that the characteristics presented in Section 1 are specific to design and make design differ from nondesign. However, we cannot refer to empirical studies that show this.

**3. Design is one, but takes different forms**

The idea that there may be different forms of design has been hinted at in informal discussions, generally without empirical or theoretical evidence (Löwgren, 1995; Ullman et al., 1988). Without any such underpinning, for example, the engineering-design methodologists Hubka and Eder (1987) assert that "the object of a design activity, what is being designed… substantially influences the design process." This assertion expresses a rather generally—more or less implicitly—accepted idea, that is, that the artefact product, characterising the design discipline (architecture, mechanical, or software design) is the variable underlying the differences we are examining here.

In "Variants in design cognition," Akin (2001) states that "in different fields of design, cognitive processes have both similarities and differences" (p. 105). The author focuses on architectural design, generally contrasting it with engineering design in his paper. Compared to other designers, architects are more inclined to use (i) "rich representations," (ii) creative, inventive strategies, (iii) non-standard problem decomposition schemata, (iv) complexity management strategies, and (v) search for alternative solutions.

(i) On the basis of his extensive research on architects, Akin (2001) affirms that these designers use many forms of both analog and symbolic, "naïve," everyday and physical, technical, and domain-specific representations.

(ii) Confronting both informal observations and experimental work (Akin, 1986a) on architects and experimental results concerning electrical engineering designers obtained by Ball et al. (1997, quoted in Akin, 2001), Akin (2001) considers as plausible that "architects tend towards creative design strategies while engineers tend to routine design." His comparison between the architects in his Akin (1986) study

    producing a richness in novel solutions to constrained, closed problems, and the engineers observed by Ball et al. (1997, quoted in Akin, 2001) generating remarkably low numbers of solutions, leads him to suppose that "these engineers tend to apply routine-design strategies even when the problem calls for a novel solution."

(iii) Akin (2001) opposes the conclusion reached by himself and by colleagues that architects adopt individual decomposition schemata, to that formulated by research colleagues (Frankenberger, 1997, and Dörner, 1997, quoted in Akin, 2001) that mechanical engineers and industrial designers use standardised schemata. This holds for both the decomposition of the global design process into design phases and that of a larger problem into smaller ones.

(iv) For Akin (2001), the way in which designers recompose a comprehensive design solution from partial ones as an indicator of the way in which they manage complexity. Based on a study he conducted in 1994 (referred to in his chapter), Akin (2001) explains how architects use ad hoc strategies to integrate partial solutions into global ones. He opposes this approach to the predetermined procedures that electronic or mechanical designers use to handle the interaction between the parts of a VLSI circuit or a mechanical assembly.

(v) Akin (1986a) observed that architects continue their search for alternative solutions even if they have already formulated a satisfactory concept. They do not commit themselves prematurely to an early selected kernel idea, something that is often considered a general characteristic of designers (cf. characteristic iv in Section 1).

Concerning the second point: various authors have observed that other designers than architects—for example, product designers (Dorst & Cross, 2001; Rodgers, Green, & McGown, 2000; Van der Lugt, 2002)—may also act in creative, flexible ways (something distinctive for architects, according to Akin).

Concerning the last point, we have requested more inspection of the basis for the conclusions about premature commitment that are generally advanced in the cognitive design research literature (Visser, 2006b). There are experimental results supporting Ball and Ormerod's (2000) hypothesis that design induces early fixation on a kernel idea when designers are working individually and/or in artificially restricted situations, and that professional designers collaborating in "natural," real work situations may come up with more alternative solutions (Visser, 1993a). There are, however, also studies (1) on teams working in de-contextualised situations that show designers willing to reconsider early concepts (Smith and Tjandra, 1998, referred to in Cross, 2001a), and (2) on individually working designers who come up with several solution ideas, (2a) while they are working in artificially restricted conditions (Eastman, 1969: bathroom design;

Fricke, 1999: engineering design; Whitefield, 1989: mechanical design), but also (2b) in natural situations (Reitman, 1964: musical composition) (see also Atman, Chimka, Bursic, & Nachtmann, 1999, quoted in Cross, 2004b). The apparent contradiction between these observations might be removed in at least two ways.

(i) Designers may aspire—or simply think or declare—to refrain from premature commitment, but in fact not put these ideas in practice (see also Malhotra et al., 1980; cf. our observation that designers' accounts about their activity often do not coincide with their actual activity, Visser, 1990).

(ii) Early on in the design process, working at a conceptual level, designers may select a kernel idea, but afterwards they may refrain from premature commitment at a more concrete or detailed level, for example, by not fixing all values for its variables.

Purcell and Gero (1996) have observed a difference between mechanical and product designers as regards their susceptibility to use features of example designs. In certain situations, mechanical designers showed "[design] fixation in the traditional sense of reproducing the characteristics of [an example] design, including incorrect features" (p. 381). They did so when "the example shown embodied principles that were typical of the knowledge base of the discipline" (p. 381). However, when they received innovative design examples, they seemed to "identify [the core innovative principle involved in the example] and then explore how this could be used in the particular design situation" (p. 381), leaving out of the designs they produced many of the specific aspects of the example. With the product designers, the fixation effect was completely absent. However, to produce innovative designs, these designers did not use the innovative examples either. Maybe they became "fixated" "on being different" (p. 381); maybe their search was for difference rather than for innovation (p. 380). The authors suggest two sources for the observed differences between the designers from the two disciplines. First, education: that of product designers may emphasize creativity and the search for many different ideas. Second, the more or less varying and/or articulated character of knowledge in a domain: "the areas of knowledge that make up industrial design are more diverse than those studied in mechanical engineering" (p. 374) and many of them "are associated with less well articulated bodies of knowledge than those that make up the knowledge base of mechanical engineering. For example, aesthetics plays a prominent role in industrial design education…, while it plays little formal role in mechanical engineering" (pp. 374-375).

Through several studies of designers from different domains working in their daily professional situation (software and various types of mechanical design), we have been able to identify differences between such professionals working on industrial or other commercial design projects and design-knowledgeable participants (generally students) solving design problems in artificially restricted situations (generally laboratory experiments) (Visser, 1995, 2006b, 2006c). Three notable differences are the following. (1) In software, mechanical, or other professional design projects, designers organise their activity in an opportunistic way, whereas in simpler, more restricted situations, designers are often able to follow systematic decompositional approaches (as generally prescribed by normative methods, e.g. top-down, breadth-first; cf. our discussion above in Section 1). (2) Reuse of elements from previous projects seems a specific professional design approach—even if it has also been observed in experimental research (Détienne, 2002). (3) In our study of a professional software designer (Visser, 1987b), we noticed that user considerations were among his guiding principles (leading him, for example, to adopt certain variable-naming strategies), an observation that has not been mentioned by researchers studying design in artificially restricted situations. These observations, which do not seem specific to a particular domain of design, may point to the influence that is exerted on the activity of design by (1) design education, (2) the complexity of a design project, and (3) the design setting.

In the cognitive design research literature, one frequently encounters allusions to, or implicit testimonies of the specific character of software design compared to other types of design (see Visser, 2006b; 2006c for a discussion of these attitudes). Even if design of HCI is much less the object of discussion in this context, researchers studying software design or HCI have themselves also contrasted their domain of research with other domains. The responsible variables remain, however, unexplored. In their bibliographic cocitation analysis, Atwood, McCain, and Williams (2002) found that a set of authors representing Software Engineering design methodologies was "essentially unconnected with the remainder of the author set" (p. 129). They noticed, "software design has its own design literature" (p. 132). On the other hand, generic design journals, such as *Design Studies*, *Design Issues*, or *The Journal of Design Research* rarely publish papers on software or HCI design. The separation between software and HCI, and other types of design holds for scientific events as well. Conferences in the domain of design research concern either software design and/or HCI (i.e., treated either together or singly), or other types of design. Of course, there are specialised conferences in many domains of design, but when they announce, as their object, "design" without any further specification, conferences generally do not expect papers on software or HCI (for a list of example references, see Visser, 2006b, Section. 22.1).

Refining our analysis initiated in Visser (2006b; 2006c) and continued in Visser (2006a), this paper proposes three dimensions that we suppose underlie differences between forms of design: the design process, the designer, and the artefact. Under each of these dimensions, we propose several variables.

3.1 PROCESS

Various process-related variables may affect designers' cognitive structures and activities, and the designs they produce. We identified the organisation of the design process, the tools used, and the place of the user in the design process, possibly specified into two or more variables that are more specific.

*3.1.1 The organisation of the design process*

The way designers plan to organise their task or the process they are involved in is liable to influence their activity. Be the organisation imposed by one's hierarchy, or devised by oneself, it works as other tools: it not only structures, but also guides people's activity, through immaterial and material means, such as design methods and other tools, be they representational, or calculation and simulation aids (cf. subsection 3.1.2, *Tools in use*).

*The time scale of the design process*. Design is considered an off-line activity. One might thus naively suppose that designers, contrary to, for example, controllers of dynamic situations, have all their time to think over their projects, to analyse and change views, to discuss and confront their opinions with colleagues. The reality is different. First, most industrial, or other professional design projects generally take place under temporal constraints. Their stringency, however, may differ depending on external organisational (due to the workshop or the client), artefactual, and other factors. Second, planning—both as a design activity in itself and as a component of other design activities (Visser, 1994b)—is obviously subject to temporal variables. Several early empirical studies have examined the role of temporal constraints in the context of planning, for example, the famous study on route-plan design by Hayes-Roth and Hayes-Roth (1979). We will come back to temporal constraints in our discussion of "designing in space versus designing in time" (section 3.3.3).

*Individual versus collective design*. Certain artefacts are designed generally by an individual designer, others are usually the work of a team. Complexity and size of the artefact (two dimensions mentioned by Gross, 2003) may play a role in this association, but are certainly not the only variables. Product design is often performed by individual designers, whereas many engineering design projects are conducted collectively—but, of course, these are only tendencies.

We have defended elsewhere that there is no reason to suppose that cooperation modifies the nature of the basic cognitive activities and operations implemented in design (i.e., generation, transformation, and evaluation of representations) (Visser, 1993a). Because cooperation proceeds through interaction, it introduces, however, specific activities and influences designers' representational structures (both on sociocognitive and emotional levels). Some examples of such activities are coordination, operative synchronisation, construction of interdesigner compatible representations, conflict resolution, and management of representations that differ between design partners through confrontation, articulation, and integration. Activities involving argumentation—that is, in our view, activities aiming to modify the representations held by one's interlocutors—obviously play a particularly important role. The construction of interdesigner compatible representations (Visser, 2006b, 2006c), their existence beside designers' private representations, and their management introduce factors that may add complexity to collective design situations compared to individual design.

### 3.1.2 Tools in use

Given our view of design as the construction of representations, we privilege representational tools in this discussion, especially concerned with external representations and the means to produce them. Designers' internal (mental) representations evidently also play a crucial role in their activity, but these representations are mainly dependent on other components of the situation and on individual factors.

*Design methods*. By definition, designers proceed differently depending on the method they follow. In the domain of software design, several authors have compared the use of different design paradigms (in the sense of design methods) and observed differences, both with respect to activities and to resulting designs.

Lee and Pennington (1994), for example, have shown these two types of differences between software design using an object-oriented and using a procedural paradigm. With respect to the activity, the differences concerned the domain and solution spaces developed, the duration of problem domain analysis and of solution evaluation. As also observed by other authors (see references in Lee & Pennington, 1994), the resulting designs "[reflected] fundamentally different models of the solution. Procedural design methodologies result in designs in which the modules represent procedures that complete subparts of the task, whereas object-oriented methodologies result in modules that represent objects in the environment" (p. 581; for other differences, see the paper).

Kim and Lerch (1992), in their comparison between object-oriented (OOD) and "traditional functional decomposition" (TFD) methodologies, expected "OOD to radically change the cognitive processes in logical design" (p. 491). Based on the preliminary results obtained in a pilot study, the authors noted that "OOD may achieve substantial time savings over TFD in logical design.... 1) by simplifying rule induction processes used in functional decomposition; 2) by guiding designers on how to build more effective problem spaces; and 3) by allowing designers to run mental simulation more efficiently and more effectively" (p. 489).

*The maturity of a domain* may influence the availability—and thus the use—of tools. In 2004, the NSF launched a "Science of Design" program aiming to "develop a set of scientific principles to guide the design of software-intensive systems" (Science of Design, 2004). An underlying idea was that "in fields more mature than computer science [such as architecture and other engineering disciplines, for example, civil or chemical engineering], design methodology has traditionally relied heavily on constructs such as languages and notational conventions, modularity principles, composition rules, methodical decision procedures and handbooks of codified experience .… However, the design of software-intensive systems is more often done using rough guidelines, intuition and experiential knowledge."

As noticed above, research in the domain of software design has shown that design methodologies may have an influence on the activity and the resulting design. One may suppose that being familiar with the constructs and other tools that have been developed in a domain may influence, probably facilitate, a designer's activity—even if cognitive design research has shown the difficulty of designers' *effectively* working according to design methodology prescriptions (Carroll & Rosson, 1985; Visser, 2003; Visser & Hoc, 1990).

One may notice that related to the idea that underlies the present variable and that is only touched upon here, is the question of well-defined versus ill-defined problems and the implications for the nature of the activities involving these problems (see Visser, 2006b, 2006c). From a cognitive-activity viewpoint, most or all ill-defined problems might be analysed as design problems (Visser, 1993b). Going one step further, Falzon (2004) proposes to adopt design as a paradigm for analysing all problem-solving activities. Eventually, Falzon posits, each design problem becomes a state-transformation problem (the type of problem typically examined in classical cognitive-psychology laboratory research), because of people's acquisition of expertise and habits, and of technological evolution. Falzon nevertheless also notes the possibility that there will always remain multiple situations in which people refuse themselves to refer to procedures and routines. As an example, he refers to a study by Lebahar concerning painters who try to establish conditions that rule out the possibility to refer to routines.

*External representations.* According to Zhang and Norman (1994), external and internal representations differentially activate perceptual and cognitive processes. With Scaife and Rogers (1996), we presume that things are less systematic, and more complex. Nevertheless, we suppose that the use of internal and that of external representations involve processing differences. Therefore, designing may differ between situations depending on the importance of certain types of representations. One may suppose that, for example, design of physical artefacts (e.g., architectural or mechanical design) differs from design of symbolic artefacts (e.g., procedures or organisations).

Indeed, one of the factors underlying the differences between software and other types of design that are often stressed may be due to the different types of representations primarily used. A classical result in cognitive psychology research on external representations concerns the influence of representation "formats" on problem solving. The standard approach to show this effect has been to compare the way in which isomorphic problems were solved (for example, the Tower of Hanoi and the Tea Ceremony). Another difference in format is that between alphanumeric and figurative representations. The possibilities provided by sketches and other types of drawings compared to those offered by purely alphanumeric representations, for example, with respect to the ease of visualisation and manipulation and their corollaries may facilitate simulation and other forms of evaluation of what are going to become physical artefacts.

This observation surely does not only apply to classical (i.e., nonvisual) forms of software design. It probably also holds for other symbolic artefacts, such as other types of procedures, plans, and organisational structures.

According to Akin (2001), architects differ from designers in other domains with respect to their relatively more frequent use of (1) analogue compared to symbolic representations and (2) varying representations. The author attributes this greater representational variety to architectural design's situated and user-dependent character. Akin also points to the lack of universally accepted representational standards in architecture (cf. other elements of Akin's position presented in the introduction of Section 3). We already put into perspective Akin's view that architectural designers are particularly resourceful and flexible. However, the lack of standards, be they universally accepted or not, may indeed be of influence (see also our remark concerning *The maturity of a domain*, Section 3.1.2).

According to Zimring and Craig (2001), disciplines of practice differ with respect to the ease or even the possibility for users to understand the intermediate and final representations of the artefact generally used in the domain. They assert that, for example, architectural drawings of layout or appearance of the building to be designed "are at least theoretically understandable by end-users…, resulting in patterns of collaboration, testing and accountability that differ significantly from those associated with more 'invisible' design processes. An engineer, for example, working on a car engine is not likely to collaborate with end users directly, given the difficulty the average car driver has in understanding the mechanics of engines." (p. 128)

With respect to the role of representation, we wish to state explicitly that the importance of its role undeniably also depends on the designer (discussed in Section 3.2).

*Possible means for evaluation*. Domains differ in the methods and other tools that may be used in order to evaluate design proposals (Malhotra et al., 1980, pp. 129-130). In engineering, more or less "objective" measures and other criteria for future artefacts' performance can be used and different proposals can be ranked rather objectively. One can calculate whether a particular design (e.g., a bridge) meets particular functional requirements, such as accommodation and maximum load. The results of qualitative evaluation used in other domains, based on subjective criteria such as aesthetics, for example, may be more difficult to translate into a "score," and thus to compare. In between the extremes of completely objective and entirely subjective evaluation, exist different types of simulation, physical and mental.

*3.1.3 The user in the design process*

Designers design for other people, the "users" of the artefact product. In each domain of design, users are central—even if not always for the designers and even if the use of artefacts may (seem to) be more or less direct (cf. also *The artefact's impact* in Section 3.3.2). Naïvely, one might think, for example, that industrial design products (such as, pens, chairs, household boxes) are in "direct" use by their users, whereas the relation between the product of a city-planning project and its users is much less direct. Yet, domains differ with respect to their common practices regarding the way in which designers usually take into account the potential, future users and their use of the artefact product.

*Integration of user data into the design*. In HCI, for example, there is a tradition and, correspondingly, considerable effort towards developing methods to integrate user data into the design. This has varied from such data being introduced into the design by design participants who "know" the users but are not users themselves, to approaches such as participatory design in which the users have themselves a voice in the design process (Carroll, 2006).

It seems likely that the number and variety of participants who take part in a design process influence this process, probably more its socio-organisational than its cognitive aspects (see also Section 3.1.1, *Individual versus collective design)*. Yet, on a cognitive level, the difficulty of integration may increase with the number of different representations to be integrated—thus with the number of types of participants. In addition, the participation of "nontechnical" design participants (what users generally are) may introduce a specific difficulty, both for the nontechnical participants themselves and for their professional design colleagues.

Gross (2003) mentions two specific user-related variables on which a piece of software is like a building: the difference or equivalence between client and user, and the type of use or user, which may change more or less, or may remain constant. These two variables get a particular weight in the context of the abovementioned influence that number and variety of participants may have on the design process.

3.2 DESIGNER: INTERINDIVIDUAL DIFFERENCES

Differences between designers may affect both their activities and their representations. This influence may occur by way of one or more of the variables proposed hereafter. The use of certain types of representations or other tools may influence design thinking, but a particular designer may be more inclined to adopt a particular type of representation, or feel more at ease with its use.

*3.2.1 Design expertise*
A classical cognitive-psychology result confirmed in cognitive design research is that experts and novices in a domain differ as to their representations and activities. Experts, for example, "recognise underlying principles, rather than focusing on surface features of problems" (Cross, 2004b, p. 432). Expertise has been examined in many experimental studies on design (Chi, Glaser, & Farr, 1988; Cross, 2004a, 2004c; "Expertise in Design," 2004; Glaser, 1986; Glaser & Chi, 1988; Reimann & Chi, 1989). In Section 2, we saw that besides novice designers differing from expert designers, design expertise in itself differs from that in other fields (Cross, 2004b).

We have proposed to distinguish, in addition to *levels* of expertise, *types* of expertise (Falzon & Visser, 1989; see also Visser & Morais, 1991). We have indeed observed how designers who were "experts" in the same domain, but had different prior experience in that domain (workshop vs. laboratory), exhibited (1) different types of knowledge and (2) different organisations of their knowledge—a result comparable to that regarding levels of expertise.

*3.2.2 Routine character of a task*

The routine character of a task is not an objective characteristic of the task, but depends on the representation that people construct of it. This task characteristic is thus dependent on interindividual differences.

Most design projects comprise both routine and nonroutine tasks. In a comparative analysis of three of our empirical design studies, we have established a link between the more or less routine character of a design project and the way in which analogies are used (at the action-execution and at the action-management levels) (Visser, 1996). This, in turn, influences, at least in part, the possibilities a designer has for reuse in a design project.

*3.2.3 Idiosyncrasies*

"The reality of professional design practice seems to be that individual designers have differing design abilities—some designers just seem to be better than others, and some are outstandingly good." (Cross, 2002a, p. 14) In addition to studies comparing experts and novices, there are clinical studies on experts that have led researchers to identify specific characteristics of particular experts (Cross, 2001c, 2002a). It seems, for example, that, contrary to most architects, Frank Lloyd Wright could imagine and develop a design entirely without using external representations, not sketching or drawing, be it until an advanced stage of the project (Tafel, 1979, quoted in Ball & Christensen, 2007), or even until the very end of the design process (Weisberg, 1993, quoted in Bilda & Gero, 2005). Cross (2002a), who mentions several studies on exceptional designers, presents three case studies he has performed himself on creative design in engineering and product design.

Concerning the three designers whom he considers "exceptional," Cross (2002a) notices that "there appear to be striking similarities in their design strategies, which suggest that general models might be constructed of design expertise and creative processes in professional design practice" (p. 14).

*3.2.4 Different "types of people"*

An idea often encountered, especially among architects themselves, is that, rather than possessing idiosyncratic characteristics, architects (as members of the architectural profession) are of a "special kind," a different "type of people" than other designers, especially, software designers or engineers. Architects would have a different "personality": they would be, for example, more creative and more aesthetics-oriented. Several researchers, architects themselves or not, also defend such an idea. Akin (2001), for example, considers that architects attribute another value to creative and unique designs than other designers do. Besides advancing more tangible differences presented above (see the introduction of Section 3), Akin (2001) claims that "the profession of architecture rewards the heart while engineering rewards the brain" (p. 105). He notes explicitly, however, that the specificities of architects are not biological or physiological, "or even fundamental intelligence related." Akin (2001) believes that "it is a matter of ethos and culture fostered in a given profession, its educational philosophy and the predisposition of its participants."

"Personality" is a familiar notion in psychology (nearly every theoretical tradition in psychology has its own personality theory), which, as many notions from the social sciences and humanities, is widely used outside of this discipline. The personality-related difference between architects and other designers, however, does have no scientific basis: we are unaware of any empirical study concerning possible "personality differences" between designers working in different domains.

3.3 ARTEFACT

We have identified three artefact variables: social embeddedness, type of artefact (instantiated by structures versus processes), and artefacts' evolution. Gross (2003) proposes the proportion of reusable components in the artefact's structure as one of the factors that make "a piece of software like a building," but we do not have any data or hypotheses concerning such a variable.

*3.3.1 Social embeddedness*

Referring to Rittel and Webber (1973/1984), Zimring and Craig (2001) point to the social embeddedness of planning and design problems. In their analysis of societal planning problems as "wicked," Rittel and Webber (1973/1984) indeed attribute the major part of this wickedness to the social embeddedness of these problems. Neither Rittel and Webber (1973/1984), nor Zimring and Craig (2001) define the notion. Our definition is based on Edmond (1999)s' constructivist approach: social embeddedness refers to the extent to which an appropriate characterisation of an agent's activities and representations requires that one include the agent's social environment ("the society of agents" in Edmond's terms) as a whole in the characterisation.

We consider that it is not an externally defined task that is more or less socially embedded, but people's representations involved in dealing with it. Qualifying a design project as socially embedded is then shorthand for qualifying as such the designer's representations of that project (an analogous remark holds for "a project's ill definedness" as shorthand for "the ill definedness of the representations that the designer has constructed of the project"). A project can become socially embedded because the designers or other stakeholders consider necessary to take into account the insertion and future position of the artefact in its environment (characterised by its users and/or the global society).

In apparent opposition with what we advanced above concerning user involvement in design, one might think that a product design project is less socially embedded than an urban project; that development of HCI involves more social embeddedness than traffic signal design, for example. Given our view that the view held by the designers or other stakeholders makes a project socially embedded, this quality is not limited to what are generally considered "social" or "societal" problems. Even if societal planning problems generally will be socially embedded, this characteristic is not specific to planning problems: for example, planning one's route through a city (Chalmé, Visser, & Denis, 2004; Hayes-Roth & Hayes-Roth, 1979) or planning a meal (Byrne, 1977) are not necessarily typical instances of socially embedded problems. Yet, planning a meal can become socially embedded depending on the meal "designers"' view of their guests and of the consequences occasioned by the meal's more or less greater "success"—and as such, designers' activity will be influenced by their view.

The influence of social embeddedness on a designer's activities is probably similar to that of ill definedness. For example, socially embedded problems probably have various, different solutions. These solutions may be considered more or less appropriate, more or less acceptable, depending on the criteria adopted by the person who judges them.

A related issue in which societal questions play an important role is the consideration of an artefact's user in the design of the artefact (see Section 3.1.3, *The user in the design process*). Winograd (1996) considers that its user-oriented character makes software design comparable to architectural and graphic design, and different from engineering design. He considers, however, that the design of interactive software is completely different from other software design (Winograd, 1997). Among the arguments advanced for these claims, none is based on cognitive analyses of the activity.

Simon (1969/1999), in *Sciences of the Artificial*, implicitly establishes a radical distinction between design activities in engineering and in social design. In his discussion of social planning, Simon (1969/1999) states that "representation problems take on new dimensions" in this form of design (and, maybe, also in inventive engineering design, see Visser, 2006b, 2006c) compared with the "relatively well-structured, middle-sized tasks" of engineering and architectural design (p. 141), which he presents—implicitly—as the prototypes of design. For "real-world problems of [the] complexity" of social planning, Simon considers that designers may refer to "weaker" criteria than in the case of standard design. Processes such as "search guided by only the most general heuristics of 'interestingness' or novelty" may provide "the most suitable model of the social design process" (Simon, 1969/1999, p. 162). With respect to the sources of social design problems' greater "complexity," Simon suggests that differences of at least three types may be involved: problems' degree of structuredness (here qualified as "definedness"), their size, and the nature of their object (see our detailed discussion of these differences in Visser, 2006b, 2006c).

It seems that only when he discusses social problems (and perhaps scientific discovery problems, see the introduction to Section 2 *Design is different from nondesign*) that Simon seriously considers human bounded rationality and takes into account the role of what he calls "representations without numbers," generative constraints such as "interestingness" or "novelty," and critical constraints such as the "defensibility" of a decision. The hypothesis that we have formulated in order to explain—at least, in part—this view of design adopted by Simon is that he considers (1) (routine) engineering and architectural design as standard design, and (2) social planning as radically different from standard design.

In ever more domains, people become convinced of the societal aspects of their action. One may suppose that this evolution will have its influence on design. "The common thread in the new approach to traffic engineering is a recognition that the way you build a road affects far more than the movement of vehicles. It determines how drivers behave on it, whether pedestrians feel safe to walk alongside it, what kinds of businesses and housing spring up along it." (McNichol, 2004) (see also tendencies such as ecodesign, ecological design, and sustainable design, see, e.g., Méhier, 2005) (cf. the next subsection, where we discuss the influence of users' interaction with artefacts on designers' activities—especially those related to the anticipation of the artefact's behaviour over time).

*3.3.2 Artefacts' evolution*

"Interactive systems are designed to have a certain behavior over time, whereas houses typically are not," according to Löwgren (1995, p. 94). Even if this assertion is questionable with respect to "behaviour" in general, behaviour over time is a variable on which artefacts differ—and the types of behaviour of different artefact products are quite diverse. An artefact's behaviour over time may be related to its impact on people (the "transformative" nature of artefacts, see Carroll, Rosson, Chin, & Koenemann, 1998), through the interaction that people engage in, and to its use by people who are not necessarily transformed by this use. It may also be due to its deterioration, dependently or independently of people. Two variables introduced by Gross (2003) are the degree to which components of the artefact may be subject to change or renewal, and the more or less extended lifetime of the artefact.

All artefacts change over time. Houses may not display "behaviour" over time, but they change. Systems such as organisations or interactive systems are subject to specific types of change. Designers are supposed to anticipate the transformation that their artefact products undergo—be it of deterioration or another evolution type. The possibility of anticipation may vary between situations (domains), not necessarily depending on the degree of impact. It depends, among others things, on the possibility to simulate the artefact, or to test it in another way. For interactive artefacts, anticipation may be performed through simulation. The future behaviour of certain technical artefacts may be anticipated based on calculations.

*The artefact's impact on people's activity and the possibility to anticipate it.* Predicting people's future use of an artefact product and further anticipating the impact of the product on human activity, is one of the "characteristic and difficult properties" of designing (Carroll, 2000, p. 39). Indeed, "design has broad impacts on people. Design problems lead to transformations in the world that alter possibilities for human activity and experience, often in ways that transcend the boundaries of the original design reasoning" (Carroll, 2000, p. 21). Gross (2003) mentions sanitary risks and safety concerns that particular uses or states of an artefact may introduce.

Even if all design has impact on people, certain domains seem more sensitive than others are. HCI, with which Carroll (2000) is especially concerned in his discussion quoted above, is an example of a domain in which design has particularly broad impacts on people. Yet, this holds for all design with societal implications.

*Distance between intermediary representations and final product.* The design of an artefact is a different activity than its implementation (Visser, 2006b, 2006c). For certain types of artefacts, however, there seems to be a relatively fluid, steady transition between the different forms that the design concept may take and the final artefact product—what may be qualified as a shorter "distance" between the two. Symbolic artefacts, such as software, are an example. This might elucidate somewhat our observation that software designers find it particularly difficult to separate design from coding (Visser, 1987b). It does not imply, however, that design and implementation are not distinct for symbolic artefacts.

It is with respect to the distance between the design concept and the final artefact product that Löwgren (1995, p. 94) opposes architectural and engineering design to "external" software design ("design of the external behavior and appearance of the product, the services it offers to users and its place in the organization").

*Delay of implementation*. Design is by definition concerned with artefact products that do not yet exist. A central aspect of designing is thus, once again, anticipation. The bases of this anticipation may vary depending on other variables (users' taking part and designer's knowledge, experience, and activities, such as simulation), but anyhow the conditions of existence, the behaviour, and the use of the artefact products will be more or less different from those anticipated: the world changes without possibility of being completely controlled.

The implementation of certain types of artefacts is much longer in coming than that of others—and not because of laziness or indifference of the workshop or the client, or due to lack of resources. Voss et al. (1983) have noticed that the solving of social-science problems is particularly difficult because of the "delay from the time a solution is proposed and accepted to when it is fully implemented" (p. 169). Such a delay clearly complicates the anticipation of the artefact's evolution and other matters involved in its evaluation (through simulation or other means). Even if this observation is particularly applicable to social-science problems, it may also hold for other types of design.

*3.3.3 Type of artefact*

"Type of artefact" may seem an evident explanatory variable for the existence of different forms of design. As noticed already, however, few elements are available concerning underlying variables. An example is software design—often considered as "essentially different" from design in other domains, but without discussion or examination of the responsible variables. One candidate variable could be the difference between structures and processes. Data concerning the influence of this variable may come from results obtained in studies concerning what may be considered particular instances of structures and processes, that is, spatial and temporal entities.

*Designing in space versus designing in time*. Studies comparing problems governed by temporal and problems governed by spatial constraints have shown that designers deal differently with these constraints (Chalmé et al., 2004; detailed in Visser, 2006b). An example of design that preferentially implements temporal constraints is planning (meal planning, see Byrne, 1977; route planning, see Chalmé et al., 2004; Hayes-Roth & Hayes-Roth, 1979). Research, however, has not yet settled clearly the specificity of the relative ease and difficulty involved in the corresponding types of design—it has even less identified the underlying factors.

Structures (which may correspond to states) are not necessarily spatially constrained, but processes have systematically temporal characteristics. By analogy to the differences between the cognitive treatment of spatial and temporal constraints, one may expect that structures and processes are represented differently (especially mentally, but also externally), thus processed differently, and therefore lead to different design activities (cf. Clancey's, 1985, distinction between configuration and planning).

## 4. Conclusion

In this section, we will first briefly review the generic-design hypothesis and then focus on the component of our augmented generic-design hypothesis that has been central in this text, that is, design *also* takes different forms.

*The validity of the generic-design hypothesis*. Given the scarcity of empirical evidence, the generic-design hypothesis needs more research. We formulated above several points of reserve concerning the only systematic study (Goel & Pirolli, 1989; 1992), but we have presented, in Sections 1 and 2, various elements of support for the hypothesis' two components. The hypothesis requires, however, more research, especially more systematic research.

*The validity of the hypothesis that design* also *takes different forms*. If because of the rare and disseminated empirical evidence, the generic-design hypothesis needs more research, this holds a fortiori for our hypothesis that design *also* takes different forms. Most variables proposed were based on a broad knowledge and analysis of the results of some 25 years of cognitive design research; they would need comparative studies about the hypothesised differences between design situations. For some of them, we were able to present precise empirical data: for differences between architectural design and engineering design (Akin, 2001), and between mechanical and product design (Purcell & Gero, 1996), between design in a professional project and design in an artificially restricted situation (Visser, 1995, 2006b, 2006c), between design activities and resulting designs in projects adopting different design methods (Kim & Lerch, 1992; Lee & Pennington, 1994); for differences due to designers' expertise (for levels of expertise, see especially Cross' work; for types of expertise, see Falzon & Visser, 1989); for the influence of design projects' routine character on the way in which designers use analogy (Visser, 1996), and the influence of idiosyncratic differences between designers (shown for certain experts, see several references in the text).

Therefore, in the analysis presented in this paper, we have introduced material that still requires further analysis, and indicated a number of directions—to be followed, modified, completed, and developed, in other research.

It is conceivable that not all variables proposed have the same degree of influence. Given our view of design as the construction of representations, we suppose that variables related to representational structures and activities are particularly influential. Referring to classical research on the influence of representation formats on problem solving, we formulated the hypothesis that sketches and other types of drawings may facilitate certain activities (such as simulation and other forms of evaluation), due to the augmented ease of visualisation and manipulation offered by such figurative external representations. Yet, variables may also depend on other underlying factors and their influence on the activity may exert itself by way of representational structures and activities.

The variables and the characteristics of the different forms of activities and cognitive structures—if their influence were to be confirmed—may have implications for design support. Given the centrality of representation in designing, the development of appropriate support modalities for representational activities and structures already suggests itself—be such modalities technological (generally, computerised) or methodological. However, according to the role of representation, and the type of representation preferentially used in specific design tasks and/or in specific design situations, the development of specific support modalities may be worthwhile. Research on these questions may take advantage of the progress already obtained in other domains, for example, those of software and HCI design. In those domains, there has been considerable research on visualisation and other visual tools, for example, on diagrammatic reasoning (see Blackwell, 1997, and the Diagrammatic reasoning site; see also the research on multiple —external— representations, e.g. by Van Someren, Reimann, Boshuizen, & De Jong, 1988).

A question that might be asked after the presentation of all these—possibly—different forms of design is: if there are so many differences between the implementations of design thinking in different situations, then what about the idea that design is a "generic" activity? In order to answer this question—and counter the underlying opposition to the generic-design hypothesis—we now come up with the fourth member of our augmented cognitively oriented generic-design hypothesis. In its complete form, we see this hypothesis as the following.

(1)     Design thinking has distinctive characteristics from other cognitive activities.

(2)     There are commonalities between the implementations of design thinking in different design situations.

(3)     There are also differences between these implementations of design thinking in different situations.

(4)     However, the commonalities between all the different forms of design thinking are sufficiently distinctive from the characteristics of other cognitive activities, to consider design a specific, generic cognitive activity.

Given the hypothetical character of the third member, which was examined here, we did not mention this fourth member before. If one defends the idea of design as a generic cognitive activity, it is, however, the counterpart of the third member. At the end of this paper, this fourth member remains completely hypothetical and requires new empirical research comparable to Goel and Pirolli's (1989) work—but preferably, in our opinion, performed in real, i.e. professional design situations.